\documentclass[structabstract]{aa}  
\usepackage{graphicx}
\usepackage{txfonts}
\usepackage{natbib}
\newcommand\T{\rule{0pt}{2.6ex}}
\newcommand\B{\rule[-1.0ex]{0pt}{0pt}}

\begin{document}
\title{Phase closure nulling of \object{HD\,59717} with AMBER/VLTI\thanks{Based on observations
    collected at the European Southern Observatory, Paranal, Chile,
    within the commissioning programme 60.A-9054(A).}}
\subtitle{Detection of the close faint companion}
\titlerunning{Phase closure nulling of HD\,59717}
\author{G. Duvert, A. Chelli, F. Malbet \and P. Kern}
\institute{Laboratoire d'Astrophysique de Grenoble and Mariotti
  Center, UMR 5571 Universit\'e Joseph Fourier/CNRS, B.P. 53, 38041
  Grenoble Cedex 9, France\\
  \email{Gilles.Duvert@obs.ujf-grenoble.fr} }

   \date{date received; date accepted}

   \abstract {}{The detection of close and faint companions is an
     essential step in many astrophysical fields, including the search
     for planetary companions. A new method called ``phase closure
     nulling'' has been proposed for the detection of such faint and
     close companions based on interferometric observations when the
     system visibility amplitude is close to zero due to the large
     diameter of the primary star. We aim at demonstrating this method
     by analyzing observations obtained on the spectroscopic binary
     HD\,59717.}{Using the AMBER/VLTI instrument in the K-band with
     $\sim1500$ spectral resolution, we record the spectrally
     dispersed closures phases of the SB1 binary HD\,59717 with a
     three-baseline combination adequate for applying phase closure
     methods. After a careful data reduction, we fit the primary
     diameter, the binary flux ratio, and the separation using the
     phase closure data.}{We detect the 5-mag fainter companion of
     HD\,59717 at a distance of 4 stellar radii from the primary. We
     determine the diameter of the primary, infer the secondary's
     spectral type and determine the masses and sizes of the stars in the
     binary system. This is one of the highest contrasts detected by
     interferometry between a companion and its parent star.}{}{}

   \keywords{stars: fundamental parameters -- stars: individual: HD~5917
     -- binaries: close -- binaries: spectroscopic --
     techniques: interferometric -- astrometry }

   \maketitle
\section{Introduction}
Since the first firm discoveries of planetary-mass companions to
stars, there has been a renewed interest in all the observational
techniques that can provide reliable estimates of binary masses and
distances. Beside the mass, the most desirable information and the
most difficult to measure is the spectrum of the companion. Imaging
the close environment of stars is a very active research field
\citep{2008Sci...322.1348M,2008Sci...322.1345K, 2009A&A...493L..21L},
and is the main motivation for several ambitious instruments proposed
recently \citep{2009AsBio...9....1C, 2008SPIE.7013E....L}, based
either on fringe nulling \citep{1978Natur.274..780B,
  1998ARA&A..36..507W} or ``extreme'' adaptive optics and coronography
\citep[see review by][]{2007prpl.conf..717B}.  These techniques are
complementary to interferometric imaging since they are blind to
regions within a few Airy disks from the central star, whereas
the interferometric observables acquired by spatially-filtered
interferometers largely come from within this Airy disk.

A large fraction of the observational effort of current optical
long-baseline interferometers is devoted to the study of the close
environment of stars. By using spatial filtering techniques, the
interferometers have gained in precision and stability. Using more
than two apertures simultaneously, together with a large spectral
coverage, near infrared beam combiner like the AMBER/VLTI instrument
\citep{2007A&A...464....1P}, can provide precise measurements
of these close environments, of less than a few hundred stellar radii. We
have recently proposed a new interferometric technique called
``phase closure nulling\footnote{Not to be confused with the
  ``Three-Telescope Closure-Phase Nulling Interferometer Concept''
  proposed by \cite{2006ApJ...645.1554D} which deals with the
  phase closure properties of \emph{nulling interferometers}. }''
(hereafter PCN), to detect and to perform the spectroscopy of faint
companions of stars \citep{2009A&A...498..321C}. This technique is
based on the modeling of spectrally dispersed phase closure
measurements of the multiple system around visibility zeros of the
primary. In these regions, there is always a spatial frequency
interval within which the phase closure signature of the companion is
\emph{larger than any systematic error} and is thus \emph{measurable}.

We illustrate the proof of concept of PCN with a
simple observational case, the bright single-lined spectroscopic
binary \object{HD\,59717}. Although the observations were
not initially intended for this purpose, and are thus very incomplete,
we can nevertheless derive the characteristics of the system in terms
of stellar diameter, separation and flux ratio. The observations and
the data reduction are presented in Sect.~\ref{sect:obsred}.
Section~\ref{sect:results} describes the derivation of the binary
parameters, from visibility and phase closure measurements. The
results are discussed in Sect.~\ref{sect:discussion}.
\section{Observations and data reduction}
\label{sect:obsred}
\subsection{Observations}
\label{sect:observations}
\begin{table*}
  \caption{Summary of observation log. Uniform disk diameters
    for calibrators were obtained with the use of \texttt{SearchCal}
    \citep{2006A&A...456..789B}.} 
  \centering
  \begin{tabular}{llccccccc}
    \hline\hline
    \T Name &Date &Time of observation  & Sp. Type &Mag.~$K$ &$\theta_{UD}$ &Fringe &Spectral coverage &DIT\\
            &     & (UTC)               &          &         & (mas)        &tracker &(nm) &(ms)\\
    \hline
    \T \object{HD\,59717} &2008-02-13  &03:03:07, 05:40:59   & K5III&-0.41&- &yes &[1925-2275] &200\\   
    \T \object{HD\,56478 } &2008-02-13 &02:26:23, 04:26:38     &K0III &4.48& $0.618\pm0.043$ &yes &[1925-2275] &200\\   
    \T \object{HD\,37350} &2008-02-13 &05:04:59    &F6Ia &2.09& $0.643\pm0.026$ &yes &[1925-2275] &200\\   
    \hline
    \T \object{HD\,59717} &2008-02-14 &03:02:50, 04:51:56, 06:11:50   &K5III &-0.41&- &no &[2110-2190] &50\\   
    \T \object{HD\,70555} &2008-02-14 &03:27:16, 05:18:42, 06:40:03   &K1III &1.53& $2.54\pm0.037$ &no &[2110-2190] &50\\  
    \T \object{HD\,45669} &2008-02-14 &03:49:44, 04:26:21, 05:48:39   &K5III &1.88& $2.19\pm0.053$ &no &[2110-2190] &50\\   
    \hline    \hline
  \end{tabular}
 \label{tab:log}
\end{table*}
The observations are part of a series of on-sky tests
performed to assess the stability of the AMBER instrument and the
corresponding accuracy of absolute calibration. They were performed
after a change of the infrared detector and the temporary removal, by our
team, of a set of polarisers that was inducing Fabry-P\'erot
fringe beating, leading to instabilities in the instrumental
visibility \citep{VLT-TRE-AMB-15830-7120}. As a consequence of this
removal, the fringe contrast decreased, but the stability improved
drastically. To characterize the stability, we observed a series of
calibrator stars of various magnitudes, at different locations in the
sky, taken from the ESO list of VLTI calibrators. One of these stars,
HD\,59717, is a triple system located at 56.3~pc, consisting of
a K5III single-lined spectroscopic binary and a G5V companion at
$22\farcs3$ distance \citep[MSC catalog,][]{1997A&AS..124...75T}. The companion was not in the field of view of
the AMBER observations, and we will refer hereafter to the SB1 close
binary as \object{HD\,59717}.  

We used
three $1.8$m auxiliary telescopes (ATs) on VLTI stations H0, D0 and
A0. In this configuration, the 3 telescopes are aligned in a very
rough East-West ($71\degr$) configuration and the ground baseline
lengths are A0-D0: 32~m, D0-H0: 64~m, A0-H0: 96~m. Due to its large
angular size, \object{HD\,59717} is well resolved in the
H0--D0--A0 configuration, the longest baseline crossing by
supersynthesis effect the first zero of the visibility curve and
providing information in the first and the second lobes.

Our set of data consists of three observations of \object{HD\,59717} performed
on 14 February 2008, using the spectral window $2110$--$2190$\,nm and
a detector integration time (DIT) of $50$\,ms. Each observation was
bracketed with those of the calibrators \object{HD\,70555} and
\object{HD\,45669}. Additionally, we make use of a single visibility
measurement taken from an observation performed the previous night 
with FINITO, the VLTI fringe tracking facility \citep{2004SPIE.5491..528G}. This
observation covers the wavelength range $1925$--$2275$\,nm, used a
DIT of $200$\,ms, and was bracketed with observations of the 
calibrators \object{HD\,56478} and \object{HD\,37350}. 
  \begin{figure}
   \includegraphics[angle=-270,width=\columnwidth]{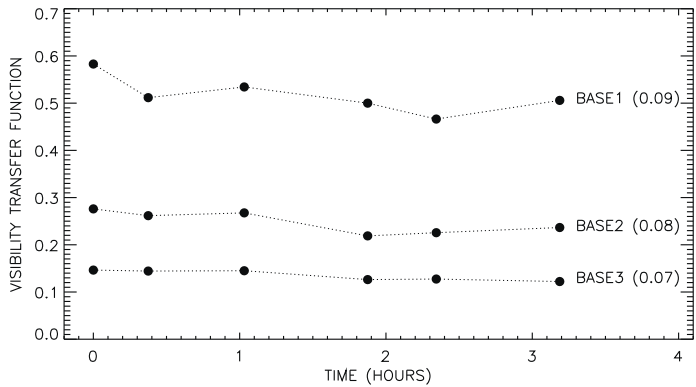}
   \caption{Transfer function computed from the 6 calibrator observations carried out on 14
     February 2008 as a function of time for the 3 baselines (shortest
     baseline at the top). The numbers in parenthesis are the relative
     visibility dispersion.}
\label{fig:calibrators}
   \end{figure}
\subsection{Data reduction}
\label{sect:datared}
Prior to the data reduction, we performed an accurate wavelength
calibration thanks to the identification of atmospheric features in
the spectrum obtained from the single observation with FINITO. We fit
the position of the atmospheric features with a second degree polynomial, 
providing a precision of a few nanometers. Then, the data were reduced
with the new AMBER data reduction method presented in \cite{paperIII}. 

Figure~\ref{fig:calibrators} shows the instrumental transfer function
averaged in wavelength, obtained for the 3 baselines as a function of
time. This transfer function results from the ratio of the measured
visibility and the intrinsic visibility of the calibrator. The number
between parenthesis is the relative visibility dispersion, computed
for each baseline, as the ratio between the dispersion of the
visibilities along the night and their mean.  This is the dominant
source error due to a residual atmospheric jitter during the
frame integration time. Other sources of error are negligible in this
respect. The intrinsic visibilities have been derived from diameters
estimated with the SearchCal software \citep{2006A&A...456..789B} and
their associated relative error is much smaller, less than $0.8$\% for
the largest calibrator on $100$m baseline.
The high spectral resolution ($1500$) prevents any spectral mismatch
between source and calibrators. Except for the single visibility data in the second lobe obtained with
FINITO (see below), the two magnitude difference between the source
and the calibrators does not matter as, in the absence of a
fringe-tracking system, the jitter is flux-independent. 
In the
spectral range $2110$--$2190$\,nm, the transfer function is a linear
function of the wavelength. We use this {\it a priori} information to
reduce the high frequency noise on the calibrator visibilities by
Fourier filtering.

The visibility data of \object{HD\,59717} obtained with the assistance
of the fringe tracker (see Table~\ref{tab:log}) are affected by a
residual, flux-dependent, jitter that will be different on our source
and its calibrators due to their magnitude difference. In principle,
we can apply a jitter correction to the data. Unfortunately, the
jitter residuals from the fringe tracking system were not included at
the time in the instrumental data. Only one visibility of high
quality, measured in the second lobe, could be calibrated with the aid
of partially redundant visibilities obtained during the second night.

The peculiarity of AMBER is to use an internal calibration
  called P2VM to estimate the coherent fluxes. Hence, the phase
  closure on a source contains the imprints of the P2VM.  Since,
  during the building process, the three beams are measured
  independently, the internal phase closure of the P2VM is not by
  construction constrained to be zero.  This is why a phase closure
  measurement with AMBER must be calibrated by substracting that of a
  calibrator. The three (spectrally dispersed) phase closures
measured just before, during and after the minimum of visibility have
been corrected by the phase closure of a calibrator star close in time. The
main error, due to the calibration process, is about $0.05$\,radian from
the internal dispersion of the calibrator phase closure measurements
along the night. To reduce the phase closure dispersion, especially at
minimum visibility, we average the bispectra (from which the phase
closure is computed), initially sampled with a frequency interval of $0.06$\,arcsec$^{-1}$, whithin a frequency interval of $1$\,arcsec$^{-1}$.
\begin{figure}
  \includegraphics[angle=270,width=\columnwidth]{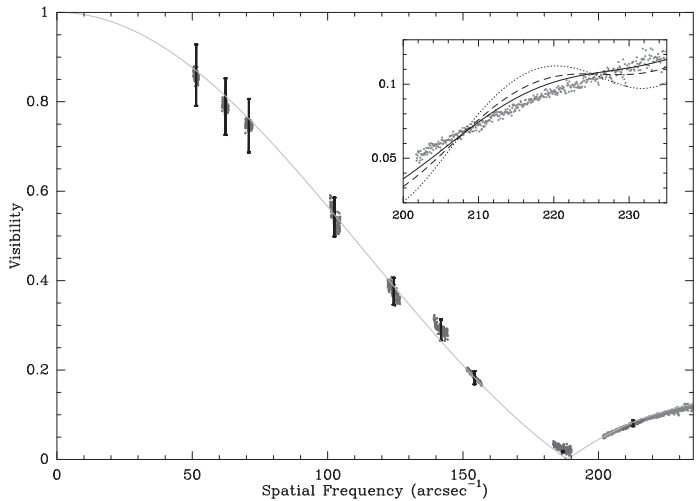}
  \caption{Visibility of \object{HD\,59717} as a function of the
    spatial frequency. Errorbars illustrate the calibration errors
    resulting from the visibility dispersion shown in
    Fig.~\ref{fig:calibrators}.  The measurements obtained with
    the fringe tracker are plotted in grey in the second lobe. The
    full curve corresponds to a single uniform disk model of diameter
    $6.436$\,mas. The insert shows the second lobe measurements and
    three binary models with separation $30$\,mas and flux ratio:
    $5\times10^{-3}$ (continuous line), $1\times10^{-2}$ (dashed line)
    , $2\times10^{-2}$ (dotted line). }
  \label{fig:sigpupv2}
\end{figure}
\section{Results}
\label{sect:results}
The calibrated visibilities of \object{HD\,59717} are shown in
Fig.~\ref{fig:sigpupv2} as a function of the spatial frequency,
defined as the ratio between the baseline and the wavelength. The
measurements are reasonably uniformly spread over the working
frequency range, with a set of measurements before, partially during,
and after the minimum visibility. The full curve represents the
best fit (excluding the points around minimum of visibility, see
Sect.~\ref{sect:results.vis2}) with a single uniform disk model of diameter
$6.436$\,mas. This is in excellent agreement with the diameter of $6.44\pm0.06$\,mas
arising from the location of the minimum of visibility around $189\pm0.02$\,arcsec$^{-1}$.
\begin{figure}
  \includegraphics[width=0.95\columnwidth]{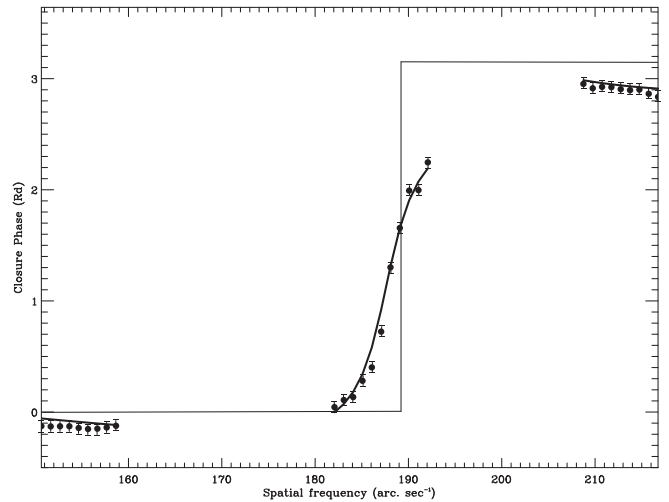}
  \caption{
Phase closure of \object{HD\,59717} as a function of
     the largest spatial frequency of the closure triangle. The thin 
    curve corresponds to the best fit for a single uniform disk model
    with diameter $6.451$\,mas. The full curve corresponds to the best
    fit with a double system and parameters: primary stellar diameter
    $6.55$\,mas, secondary projected distance $-11.2$\,mas and flux ratio
    $0.017$. 
}
  \label{fig:sigpupclosure}
\end{figure}

The phase closures are shown in
Fig.~\ref{fig:sigpupclosure}. The transition between the $0$ and $\pi$
values at the zero visibility crossing is smooth, markedly different
from the expected abrupt step arising from a centrally symmetric flux
distribution only \citep[see][]{2009A&A...498..321C}.

This departure from a centrally-symmetric object is expected in
a spectroscopic binary, and we can use these observations to retrieve
the geometric characteristics of the system. To do this, we fit
independently the visibility and the closure phase data with a simple
model formed by an extended uniform disk and a point source.  We did
not try to combine visibility and closure phase data because the
weight to give to each set of data is quite uncertain. Our
observations being basically east-west, we cannot derive the position
angle of the system, but only the projected distance along the
direction of observation.
As we have used the spectral dimension to
increase the frequency coverage, we assume that the object is
achromatic. Hence our model is described by 3 parameters: the stellar
diameter, the distance of the companion and the flux ratio.
\subsection{Visibility data}
\label{sect:results.vis2}
To estimate the system parameters, we minimize a standard $\chi^2$
defined as the distance between the visibilities of the model and
those of the data, weighted by the errors shown in
Fig.~\ref{fig:calibrators}. We exclude from the fit the set of points
around minimum visibility as they may easily be biased. Indeed, at
this location, the bias on the squared amplitude of the coherent flux
is twenty times larger than the useful (debiased) signal. Given the
relative error of $0.07$ on our longest baseline, an imprecision of
$0.07/\sqrt{20}=1.5\,10^{-2}$ would produce a bias ``error''
equivalent to the statistical error. Since we are not guaranteed such a
precision in the bias removal, it is better to exclude this set of
points, which otherwise would constrain nearly by itself the
output of the fit.  This restriction does not apply if a single
stellar diameter is deduced from the \emph{position} of the minima of
visibility, since the location of the minima are much less biased.

\begin{figure}
  \includegraphics[angle=-270,width=\columnwidth]{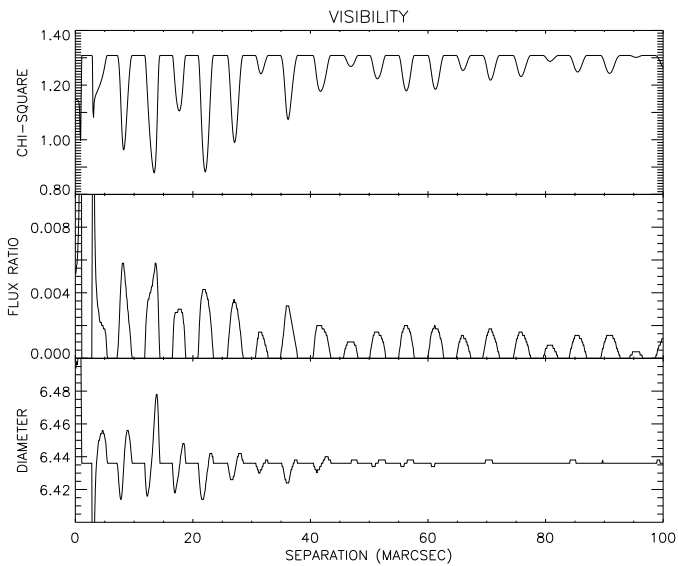}
  \caption{Best model parameters from visibility fit with an extended
    uniform disk and a point source (see Sect.~\ref{sect:results.vis2}
    for details). From top to bottom as a function of the separation:
    minimum $\chi^2$, best flux ratio and best stellar diameter.}
  \label{figure4}
\end{figure}

A rapid study of the problem shows that the $\chi^2$ has minima as a
function of the distance with a peudo-period $p \approx f^{-1}_{min}$,
where $f^{-1}_{min}$ is the frequency of minimum visibility. Hence to
produce comprehensible output, we vary the separation from $0$ to
$0.1$\,arcsec, and for each separation, we perform a fit with two
parameters, the stellar diameter and the flux ratio. The minimum
reduced $\chi^2$ (hereafter $\chi^2$), the best stellar diameters and
flux ratio are displayed in Fig.~\ref{figure4} as a function of the
separation.  

The $\chi^2$ presents shallow minima between $1.3$ (single source) and
  $0.9$ (double source) at a set of roughly regularly
  separations. However, given that below 1\% precision,
the calibration biases on the visibility become dominant, the presence
of these minima alone does not prove the binary nature of HD~59717. In
addition, their location must be taken with extreme caution.
If we except the singular point around $3$\,mas inside the stellar
disc, the range of possible stellar diameters for all the separations
is $6.45\pm0.03$\,mas. This diameter range agrees with the
  output of the fit with a single uniform disk model of $6.436$\,mas
  (see Fig.~\ref{fig:sigpupv2}).  The possible flux ratio ranges from
$10^{-3}$ to $6\times 10^{-3}$ for all separations and, the smaller the
separation the higher the flux ratio.
\subsection{Phase closure data}
\label{sec:closure}
The best fit of the
phase closure systems with a single uniform disk model provides a
diameter of $6.451$\,mas in excellent agreement with the value of
$6.436$\,mas derived from visibility data.
\begin{figure}
  \includegraphics[angle=-270,width=\columnwidth]{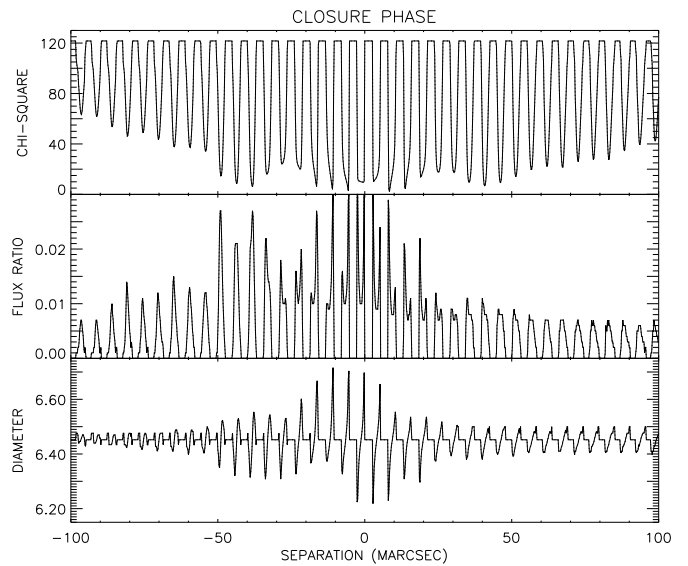}
  \caption{Best model parameters from the phase closure fit with an
    extended uniform disk and a point source (see
    Sect.~\ref{sec:closure} for details). From top to bottom as a
    function of the separation: minimum $\chi^2$, best flux ratio and
    best stellar diameter. The curves represent the best fit
    with a restricted stellar diameter in the range
    $6.42$--$6.48$\,mas imposed by visibility modelling.}
  \label{figure5}
\end{figure}
The minimum $\chi^2$ of a two-component model, the best stellar
diameters and flux ratio are displayed in Fig.~\ref{figure5} as a
function of the separation. This minimum $\chi^2$ varies from $120$
(single source) to $2.3$ (double source). 

\begin{figure}
  \includegraphics[width=\columnwidth]{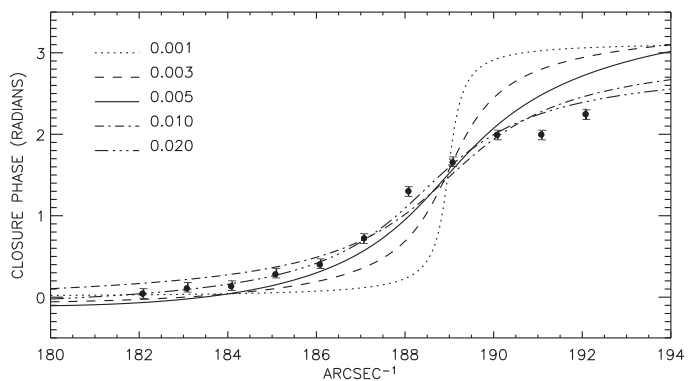}
      \caption{Best fit of the phase closure for different flux
      ratios showing that the slope of
      the transition cannot be reproduced with a flux ratio smaller
      than $5\times 10^{-3}$.}
\label{fig:model_closure_bestfit}
\end{figure}

The stellar diameter
ranges between $6.2$ and $6.7$\,mas. It is not well constrained because
we do not have the full transition, but only pieces of it.
The flux ratio varies between $6\times 10^{-3}$ and $2.7\times
10^{-2}$, which corresponds to  a
$5^{+0.55}_{-0.75}$\,mag difference between the primary and the secondary.  The slope of the phase closure is
extremely sensitive on the flux ratio. To illustrate this
effect, we plotted in Fig.~\ref{fig:model_closure_bestfit} the phase
closure together with the best fit for various flux ratios. One sees
that all flux ratios $\leq5\times 10^{-3}$ cannot reproduce the slope
of the transition.

The $\chi^2$ (Fig.~\ref{figure5}) exhibits regularly spaced and very
deep minima as a function of the separation. The minimum flux ratio of
$5\times 10^{-3}$ obtained in the fits renders separations larger than
$30$\,mas irrelevant, as they would produce oscillations that are not
seen in the visibility, especially in the second lobe (see insert in
Fig.~\ref{fig:sigpupv2}). Hence, the possible projected separations left are
$-5.5$, $+8.5$, $-11$, $+14$ and $-17$\,mas, with an error of
$\sim1$\,mas, from the width at mid-height of the  $\chi^2$ dips. 

In summary, visibility data put strong constraints on the stellar diameter,
$6.45\pm0.03$\,mas. Phase closure data do not constrain the stellar
diameter but support the binary nature of HD\,59717 (given the
impossibility of reproducing with a single uniform disk model the slope
of the phase closure around the minimum of visibility), providing a
range of acceptable flux ratios and five possible separations. 
\begin{figure}
  \includegraphics[angle=270,width=\columnwidth]{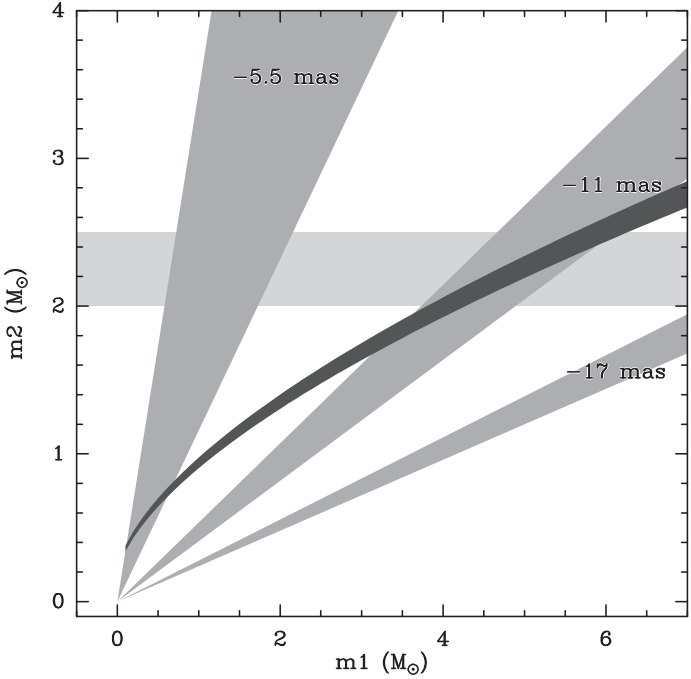}
  \caption{Loci of the solutions for the equation of mass (black),
    as a function of the mass $m_1$ of the primary and $m_2$ of the
    secondary. The regions compatible with the set of 3 separations
    given by our fit are overlayed in gray. The region
    compatible with our independent measurement of the secondary's
    flux is overlayed in light gray. }
  \label{fig:solvemass}
\end{figure}
\begin{table*}
 \caption{Literature data for \object{HD\,59717}
  \citep[from][]{2005A&A...442..365J,1918LicOB...9..116W,1997ESASP1200.....P}.} 
  \begin{tabular}{cccccccccc}
\hline\hline\T
    \B$a_0$ & $e$  & i       & $\omega_1$   & $\Omega$  &$T_0$& P &dist &$\mu$ &phot. var.\\
    (mas) &      &($\degr$)&($\degr$)     &($\degr$)  &(JD)&(d)&(pc)&($\mathrm{M}_{\sun}$)&(mmag)\\      
    \hline
    \B \T  $8.32\pm0.32$ & $0.17$ & $65.6\pm3.3$ & $349.3$ &
    $0.0\pm5.2$ &2451354.6& 257.8 &56.36 &0.164&30\\
    \hline
  \end{tabular}
  \label{tab:sigpup}
\end{table*}
\section{Discussion}
\label{sect:discussion}
The phase closure transition of \object{HD\,59717} is markedly
different from that of a single stellar disk.  Our modeling of this
phase closure transition is naturally based on the \emph{a priori}
information that the system is a binary. We do not consider here the
influence of large stellar spots because K giants are devoid of such
features \citep{1999AJ....117.1864C}, and because the larger $\chi^2$
of solutions inside the stellar radius in Fig.~\ref{figure5} make it
unlikely. Although we detect the faint companion by its effect on the
phase closure, the very incomplete information we have on its shape
around the first null \citep[compare our Fig.~\ref{fig:sigpupclosure} with
Fig.~3 of ][]{2009A&A...498..321C} prevents a precise determination
of the secondary's position using the interferometric measurements
alone. It is thus important to check whether our results are in
agreement with the already known parameters (star diameter,
spectroscopic orbit) of HD\,59717.

Table~\ref{tab:sigpup} summarizes the literature data for
\object{HD\,59717} relevant to this discussion.  HD\,59717, at a
distance of $56.36$\,pc, has a spectroscopic period of $257\fd8$ and a
reduced mass $\mu=0.164$\,$\mathrm{M}_{\sun}$
\citep{1918LicOB...9..116W}. The parameters of the photometric orbit
of HD\,59717, in particular the inclination $i=65\fdg6\pm3\fdg3$ and
semimajor axis $a_0=8.32\pm0.32$\,mas, have been obtained by
\cite{2005A&A...442..365J} from Hipparcos IAD measurements
\citep{1998A&AS..130..157V}. HD\,59717 exhibits a small amplitude
($\sim30$\,millimag in V) intrinsic variability characteristic of
rotating ellipsoidal variables
\citep{1997ESASP1200.....P,2008OEJV...83....1O}.
\subsection{Stellar diameter of the primary}
The stellar diameter of \object{HD\,59717} has been estimated by
\cite{1999AJ....117.1864C} who derive a limb-darkened (LD) angular
disk diameter of $6.86\pm0.1$\,mas. Using the relation between the LD
and UD diameters of \cite{1974MNRAS.167..475H} and a linear
limb-darkening coefficient at $2.2{\mu}$m of $0.379$
\citep{1995A&AS..114..247C}, this translates into an equivalent UD
diameter in the K band (UDK) of $6.66\pm0.1$\,mas. The
\texttt{SearchCal} web service \citep{2006A&A...456..789B} gives
UDK=$6.16\pm0.4$\,mas.  Our observations show that the diameter of the
primary is well constrained by the visibility data alone to
UDK=$6.45\pm0.03$\,mas.
\subsection{Nature of the companion}
In stellar systems for
which all the orbital and projection parameters are measured, the
knowledge of the mass $m_1$ of the main component is sufficient to
deduce the mass, hence the spectral type, magnitude, and true orbit of
the unseen companion star.  Given $m_1$, the reduced mass $\mu$, and
the inclination $i$, one solves the equation of mass:
$\frac{m_2^3\sin^3{i}}{(m_1+m_2)^2}=\mu$ for the mass $m_2$ of the
secondary. When the primary is a main-sequence star, $m_1$ is reliably
estimated using a standard mass-luminosity function
\citep[e.g.,][]{1972Ap&SS..17..134M}. Unfortunately, there is no such
thing as a unique mass-luminosity function for giant stars (as is the
case for HD\,59717), so $m_1$ is unknown and individual masses in a
binary system with a giant component cannot be derived reliably by
this method. 

Conversely, if the \emph{companion} of the giant star is
a main-sequence star, and if its luminosity can be measured, then
$m_2$ is known and the mass $m_1$ of the giant can be measured.
\begin{table*}
  \caption{Known (bold text), measured (italic) and tabulated parameters for the A and B
    components of the  \object{HD\,59717} binary.}
\centering
   \begin{tabular}{lcccccccccccc}
     \hline\hline
     \T Star & mK & MK &$\Delta$Mk & MV&$\Delta$Mv & Sp.Type & Mass &
     \multicolumn{2}{c}{Radius} & $\log{g/g_{\sun}}$ &$\log{\rho/\rho_{\sun}}$\\
      & (mag) & (mag) & (mag) & (mag) & (mag) &  &(M$_{\sun}$) &
      (mas)&(R$_{\sun}$) & &   \\
     \hline
     \T A &\textbf{ -0.41} & \textbf{-4.16} & - & \textbf{-0.55} & - &
     \textbf{K5III} & $5\pm0.5$ & \emph{3.23$\pm$0.015} & \emph{39} & -2.5 & -4.7\\
     \hline
     \T B & 5.14  &  1.39 &\emph{5.5} & 1.8  & 2.4 & A4V    & 2.0 & 0.15 & 1.8 &- &-\\ 
     \B B & 4.89  &  1.14 &\emph{5.3} & 1.3  & 1.9 & A2V    & 2.2& 0.16& 2.0  &-&- \\ 
     \B B & 4.59  &  0.84 &\emph{5.0} & 0.65 & 1.2 & A0V    & 2.5& 0.2 & 2.4 & -&-\\ 
     \hline
   \end{tabular}
   \label{tab:photometries}
  \end{table*}
  In the present case, our observations provide two new pieces of
  information: the luminosity of the secondary via our measurement of
  the flux ratio, and a choice of projections of the separation on the
  E-W direction.  Each of these pieces of information can be used
  independently to derive the mass of the giant primary, and should
  provide compatible results.
\subsubsection{Position constraints}
We first combine the constraints due to the measured projected
separation and the equation of mass.  In Fig.~\ref{fig:solvemass} we
show the loci of the solutions for the equation of mass with
$i=65\fdg6\pm3\fdg3$, $\mu=0.164$\,M$_{\sun}$ in black. At the time of
our observations, the position angle of the binary was
$\theta=-76\fdg0\pm1\fdg5$, and the projection of the separation
$sp=\alpha(1+\frac{m_1}{m_2})$, where $\alpha=-3.5\pm0.2$ is fixed by
the orbit geometry. The positive projected separations ($+8.5$ and
$+14$\,mas) are forbidden by the value of $\alpha$. The three possible
($m1,m2$) solutions left are for the projected separations $-5.5$,
$-11$ and $-17$\,mas, and are plotted in Fig.~\ref{fig:solvemass} in
gray. Their width corresponds to
the 
uncertainty ($\pm1$\,mas) on the separation.

The solutions for the binary masses lie at the intersection
between these areas and the curve given by the equation of mass. The
$-5.5$\,mas and $-17$\,mas regions give unrealistic values for the
mass distribution in the binary: the former since the primary would be
less massive than the secondary, the latter because it gives a too
high mass for the primary. This leaves only the $-11$\,mas solution,
and our mass estimate by this method is then a $5\mathrm{M}_{\sun}$
primary and a $2.2\mathrm{M}_{\sun}$ secondary.

\subsubsection{Photometry constraints}
If we use now the magnitude of the secondary derived from our observed
flux ratio, its spectral type is A0V--A2V (see
Table~\ref{tab:photometries}). Then $m_2$ lies between 2.0\,M$_{\sun}$
and 2.5\,M$_{\sun}$ \citep{1972Ap&SS..17..134M}. In
Figure~\ref{fig:solvemass} we have also reproduced this secondary's
mass range as a horizontal area shaded in light gray. It intersects
the curve of the equation of mass at $4.5<m_1<5.5$\,M$_{\sun}$, and
thus independently gives the same result as our positional measurement.

Table~\ref{tab:photometries} collects all the parameters on the
primary that can be deduced from our size and mass measurements.
Three possible identifications are proposed for the B companion,
allowing for the uncertainty on the flux
ratio. The sectral types and radii of the companion are from
\cite{1982aabs.book.....S}, masses from the mass-luminosity relation
of \cite{1972Ap&SS..17..134M}. We note that the radius and the mass of
HD\,59717 are larger than the values ($R_{K5III}=25$\,R$_{\sun}$ and
$M_{K5III}=1.2$\,M$_{\sun}$) published by
\cite{1982aabs.book.....S}. However, the masses we obtain are in
agreement with the values ($M_{Aa}=5$\,M$_{\sun}$,
$M_{Ab}>1.9$\,M$_{\sun}$) quoted in the MSC catalog
\citep{1997A&AS..124...75T} for the Aab component of HD\,59717.
\subsection{The photometric variability of HD\,59717}
The separation of the components, their size, and the
  inclination of the orbit make it impossible for the secondary to be
  eclipsed by the primary. With a $5$\,M$_{\sun}$ primary and a
  $2.2$\,M$_{\sun}$ companion, the Roche lobe $R_L$ radius of the
  primary is $R_L\approx0.45\times{d}$, where $d$ is the true distance
  of the two stars. Due to the excentricity $e=0.17$ of the companion,
  this distance varies from $23$\,mas (periastron) to $32$\,mas
  (apoastron).  With its $3.23$\,mas radius, the primary thus
  occupies between $22$\% and $32$\% of its Roche lobe, and is
  slightly ellipsoidal. This is accordance with its classification as
  a rotating ellipsoidal variable.

If the photosphere of the star was of uniform brightness and shaped by
the Roche potential, the maximum prolateness of the star would occur
at periastron and be $a/b=1.0022$.  At the time of our observations,
the deviation of the projected shape of the star from a circle would
be $\sim6\times10^{-4}$, in accordance with our hypothesis that the
primary can be modeled as a uniform disk. However, with such a small
prolateness, the change in the projected surface of the star with time
would be $<10^{-3}$, which cannot account for the $\sim30$\,millimag
photometric variability, which requires $a/b\sim1.02$. The variability
of HD\,59717 could be due to a difference of surface brightness
between the hemisphere of the star facing the companion and the other
hemisphere. Although this effect should, to a first approximation,
translate as a displacement of the photocenter of the star and would
not change the phase closure values, it will be investigated more
closely in a future study aimed at confirming at another epoch, with
dedicated observations, the detection of HD\,59717 companion by phase
closure nulling.
\section{Conclusion}
We have shown that, by fitting a simple model on the
spectrally-dispersed phase closure measurements made by AMBER on the
bright SB1 binary \object{HD\,59717}, observed near the first zero of
the visibility distribution, we can resolve the close ($\sim4$ stellar
radii at the time of observation) pair with a five magnitude difference in
brightness. This result, obtained with only 15 minutes of on-sky
integration and $1.8$\,m apertures, is two magnitudes fainter than
reported detection limits \citep[applying Eq.~A.2 of][for
$\rho=11$\,mas]{2006A&A...450..681T} and is an illustration of the
potential of the phase closure nulling method proposed by
\cite{2009A&A...498..321C}.  We have detailed the specific data
processing involved, derived the spectral type of the companion and
measured the individual masses and sizes of the system
(Table~\ref{tab:photometries}).

The PCN method employed here is of interest for the characterisation
of all faint point-like sources in the immediate vicinity of stars, if
the flux from the \emph{primary} is sufficient and with the quite
restrictive condition that the primary can be fully resolved by the
interferometer.This technique potentially gives access to masses,
even spectra, of brown dwarfs and exoplanets
\citep{2009A&A...498..321C}.  In this paper we have used published
astrometric data, since our observations were of incomplete coverage
both in spatial frequency and position angle, but dedicated PCN
observations taken at different epochs should suffice to characterize
companions and fit their orbit. For exoplanets detected by radial
velocity techniques, a single PCN measurement of the secondary's
position will solve for inclination of the orbit of the planet.
Moreover, since radial velocity techniques can be confused by the
presence of stellar spots \citep[see][]{2008A&A...489L...9H}, PCN can
provide the independent detection needed for confirmation of the
presence of an exoplanet.
\begin{acknowledgements}
  We warmly thank H. Beust and X. Delfosse for their help in the
  arcane fields of stellar photometry and binary orbits.
  We are greatly indebted to the anonymous referee who made very useful
  comments on an earlier version of this paper.

  This research has made use of the \texttt{SearchCal} and
  \texttt{ASPRO} services of the
  Jean-Mariotti Centre\footnote{Available at http://jmmc.fr}, of CDS
  Astronomical Databases SIMBAD and VIZIER, of NASA Astrophysics Data
  System Abstract Service program.
\end{acknowledgements}
\bibliographystyle{aa} 
\bibliography{11037} 
\end{document}